\documentclass[aps,pre,nofootinbib,amsmath]{revtex4}
\usepackage{graphicx}
\usepackage{amssymb}
\usepackage{bm}

\begin{document}
\title{Short and Long Range Screening of Optical Singularities}
\author{David A. Kessler}
\affiliation{Department of Physics, Bar-Ilan University, Ramat-Gan 52900, Israel}
\author{Isaac Freund}
\affiliation{Department of Physics, Bar-Ilan University, Ramat-Gan 52900, Israel}

\begin{abstract}
Screening of topological charges (singularities) is discussed for paraxial
optical fields with short and with long range correlations. \ For short
range screening the charge variance $\left\langle Q^{2}\right\rangle $ in a
circular region $A$ with radius $R$ grows linearly with $R$, instead of with 
$R^{2}$ as expected in the absence of screening; for long range screening $%
\left\langle Q^{2}\right\rangle $ grows faster than $R$: for a field whose
autocorrelation function is the zero order Bessel function $J_{0}$, $%
\left\langle Q^{2}\right\rangle \sim R\ln R$. \ A $J_{0}$ correlation
function is not attainable in practice, but we show how to generate an
optical field whose correlation function closely approximates this form;
screening in such a field is well described by our theoretical results for $%
J_{0}$. \ $\left\langle Q^{2}\right\rangle $ can be measured by counting
positive and negative singularities inside $A$, or more easily by counting
signed zero crossings on the perimeter $P$ of $A$. \ For the first method $%
\left\langle Q^{2}\right\rangle $ is calculated by integration over the
charge correlation function $C\left( r\right) $, for the second by
integration over the zero crossing correlation function $\Gamma \left(
r\right) $. \ Using the explicit forms of $C\left( r\right) $ and of $\Gamma
\left( r\right) $ we show that both methods of calculation yield the same
result. \ We show that for short range screening the zero crossings can be
counted along a straight line whose length equals $P$, but that for long
range screening this simplification no longer holds. \ We also show that for
realizable optical fields, for sufficiently small $R$, $\left\langle
Q^{2}\right\rangle \sim R^{2}$, whereas for sufficiently large $R$, $%
\left\langle Q^{2}\right\rangle \sim R.$ \ These universal laws are
applicable to both short and pseudo-long range correlation functions.
\end{abstract}
\maketitle

\section{ INTRODUCTION}

Random (and other) paraxial optical fields generically contain numerous
point topological singularities (defects) in a plane (the $xy$-plane)
oriented perpendicular to the propagation direction ($z$-axis). \ These
singularities, which include phase singularities (optical vortices) \cite{NyeBer74,Ber78}
polarization singularities (C points)~\cite{Nye99}, gradient singularities (maxima,
minima, and saddle points)~\cite{Str94}, and curvature singularities (umbilic
points)~\cite{BerHan77}, are the defining features of the optical field, and are
characterized by signed winding numbers (topological charges) $q_{\pm }$. \
Generically, vortices and gradient singularities have charge $q_{\pm }=\pm 1$%
, whereas the charge of C points and umbilic points is $q_{\pm }=\pm 1/2$.

Like electrostatic charges, topological charges screen one another: positive
charges tend to be surrounded by a net excess of negative charge, and vice
versa~\cite%
{Hal81,LiuMaz92,RobBod96,FreWil98,BerDen00,Den02,FreSos02,%
Fol03,Den03,Wil04,FolGnu04,FreEgo07,EgoSos07}. A formal measure of screening is the charge correlation
function $C({\bf r})$~\cite{Hal81}, which by convention is constructed to measure
the net excess of negative charge surrounding a positive charge at the
origin. \ When 
\begin{equation}
\iint_{-\infty }^{\infty }C({\bm r})d{\bm r}=-1\quad \left( -\frac{1}{2}
\text{ for C and umbilic points}\right)  \label{Eq. 1}
\end{equation}
screening is said to be complete, whereas for partial screening $-1<
\iint_{-\infty }^{\infty }C({\bf r})d{\bm r}<0$. \ Here we concern ourselves
with complete screening.

In generic optical fields the average net charge $\left\langle
Q\right\rangle =0$ , but there are large fluctuations: these are
characterized in lowest order by their variance $\left\langle
Q^{2}\right\rangle $. \ For a purely random collection of $N$ singularities
with say charges $q_{\pm }=\pm 1$, $\left\langle Q^{2}\right\rangle \sim N$.
\ Screening significantly reduces these fluctuations, to an extent which
depends on the rate of fall-off of $C(r)$ for large $r$. \ In the generic
case where $C(r)$ decays sufficiently rapidly, canonically exponentially, $%
\left\langle Q^{2}\right\rangle \sim N^{1/2}$~\cite{FreWil98}; this is the hallmark of
short range screening,. \ For a slowly decaying $C(r)$ however, we shall see
that the charge variance grows faster than $N^{1/2}$; we call this regime
long range screening.

$\left\langle Q^{2}\right\rangle $ and $C({\bf r})$ are closely related, and
given the latter the former can, in principle, be calculated. \ Here, we
explore in detail the behavior of $\left\langle Q^{2}\right\rangle $ and its
relationship to $C({\bf r})$ for long range screening. \ Specifically, we
examine a particular example, first suggested in  \cite{BerDen00}, and obtain
analytical results for the behavior of $\left\langle Q^{2}\right\rangle $ in
a large circular region of radius $R$, finding $\left\langle
Q^{2}\right\rangle \sim R\ln R$.

In elliptically polarized paraxial fields C points are vortices (zeros) of
the right and left handed circularly polarized components of the field~\cite{Nye99}, so our results are applicable also to these singularities. \ $C({\bf r})$
is also known for the stationary and the umbilic points of random circular
Gaussian fields, such as the real and imaginary parts of the optical field~\cite{Wil04}, but the correlation functions of these singularities differ from that
of the vortices and our results are not necessarily applicable to these
singularities.

Measurement of $C({\bf r})$ (either from experiment or computer simulation)
requires locating and characterizing all vortices in each of a large number
of independent realizations - this is a challenging task. \ On the other
hand, $Q$, and therefore $Q^{2}$, can, using the index theorem~\cite{Str94}, be
obtained from measurements made only on the boundary of the region of
interest. \ As discussed in \cite{FreWil98}, a convenient way of accomplishing this is
to count signed zero crossings of either the real or imaginary parts of the
optical field. \ Correlations between these zero crossings are described by
the zero crossing correlation function $\Gamma \left( r\right) $. \ In \cite{FreWil98}
it was argued that like $C({\bf r})$, also $\Gamma \left( r\right) $ can be
used to calculate $\left\langle Q^{2}\right\rangle $; this argument was
supported by numerical results and computer simulations, but was not
demonstrated analytically.

The plan of this paper is as follows: \ In Section II we discuss the
connection between $\left\langle Q^{2}\right\rangle $ and $C({\bf r})$ in a
natural, i.e. sharply bounded, region of finite area - first reviewing known
results for short range screening, and then presenting what are to our
knowledge the first explicit results for such boundaries for long range
screening \footnote{%
A long range screening result for a {\em Gauss-smoothed} boundary is given
in \cite{BerDen00}.}. \ These latter are the major contribution of this report. \ In
Section III we discuss the connection between $\left\langle
Q^{2}\right\rangle $ and the zero crossing correlation function $\Gamma
\left( r\right) $, showing, analytically for the first time, the equivalence
of the two seemingly different approaches to $\left\langle
Q^{2}\right\rangle $: one based on $C({\bf r})$; the other on $\Gamma \left(
r\right) $. \ In Section IV we consider a practical realization of an
optical field with long range correlations that could be used to compare
experiments with the results presented here. \ In the discussion in Section
V we consider asymptotic forms of $\left\langle Q^{2}\right\rangle $ for
very small, and very large, $R$, for physically real optical fields. \ We
summarize our main findings in the concluding Section VI.

\section{CHARGE VARIANCE IN A BOUNDED REGION AND THE CHARGE CORRELATION
FUNCTION $C\left( r\right) $}

In \cite{FreWil98} the following quantitative relationship between the charge variance 
$\left\langle Q^{2}\right\rangle $ and the charge correlation function $C(%
{\bf r})$ was derived for a bounded region of area $A$ (\cite{FreWil98}, Eq. (37)) with $%
\left\langle \left( \Delta N\right) ^{2}\right\rangle =\left\langle
Q^{2}\right\rangle $]: 
\begin{equation}
\left\langle Q^{2}\right\rangle =\left\langle N\right\rangle +\eta 
\iint_{-\infty }^{\infty }{\cal A}\left( {\bf r}\right) C({\bf r})d{\bf r}.
\label{Eq. 2}
\end{equation}
Here and throughout $\left\langle ...\right\rangle $ represents an ensemble
average, $N$ is the number of singularities with charges $\pm 1$ contained
in $A$, $\eta $ is the average number density of singularities, and ${\cal A}%
\left( {\bf r}\right) $ is the area of overlap between $A$ and its replica
displaced by ${\bf r}$.

In what follows we assume isotropy, and therefore take $A$ to be a circular
area with radius $R$. \ Elementary geometry yields 
\begin{subequations}
\begin{eqnarray}
{\cal A}\left( {\bf r}\right) &=&{\cal A}\left( r\right) =\pi R^{2}+B\left(
r\right) ,  \label{Eq. 3a} \\
B\left( r\right) &=&-2R^{2}\left( \arcsin \left( \frac{r}{2R}\right) +\frac{r%
}{2R}\sqrt{1-\left( \frac{r}{2R}\right) ^{2}}\right) ,\quad 0\leq r\leq 2R, 
\nonumber \\
&=&0,\quad r>2R.  \label{Eq. 3b}
\end{eqnarray}
Noting that $\left\langle N\right\rangle =\eta \pi R^{2}$, Eq. (\ref{Eq. 2})
becomes

\end{subequations}
\begin{equation}
\left\langle Q^{2}\right\rangle =\pi \eta R^{2}\left[ 1+2\pi
\int_{0}^{2R}rC(r)dr\right] +2\pi \eta \int_{0}^{2R}rB(r)C(r)dr.
\label{Eq. 4}
\end{equation}
Eq. (\ref{Eq. 4}) is our basic starting point. \ We proceed to evaluate it
for large $R$: first for a characteristic field with short range
correlations; then for one whose correlations are long range.

\subsection{Charge correlation function}

Halperin~\cite{Hal81} first showed that assuming stationarity and circular Gaussian
statistics, the charge correlation function for optical vortices can be
written as 
\begin{equation}
C(r)=-\frac{W^{\prime }(r)\left[ W(r)(W^{\prime }(r))^{2}+W^{\prime \prime
}(r)(1-W^{2}(r))\right] }{\pi W^{\prime \prime }(0)r(1-W^{2}(r))^{2}},
\label{Eq. 5}
\end{equation}
where 
\begin{equation}
W(r)=\left\langle E^{\ast }\!(0)E(r)\right\rangle /\left\langle \left|
E(0)\right| ^{2}\right\rangle ,  \label{Eq. 6}
\end{equation}
is the normalized autocorrelation function of the optical field $E$, and $%
W^{\prime }(r)=dW(r)/dr$, $W^{\prime \prime }(r)=d^{2}W(r)/dr^{2}$. \ Liu
and Mazenko~\cite{LiuMaz92}, and Berry and Dennis~\cite{BerDen00}, subsequently noted that Eq. (%
\ref{Eq. 5}) can be cast into the convenient form 
\begin{equation}
C(r)=\frac{-1}{2\pi rW^{\prime \prime }(0)}\frac{d\Omega ^{2}\left( r\right) 
}{dr},  \label{Eq. 7}
\end{equation}
where 
\begin{equation}
\Omega (r)=\frac{W^{\prime }(r)}{\sqrt{1-W^{2}(r)}}.  \label{Eq. Omega}
\end{equation}
The number density of vortices $\eta $ also depends on $W(r)$, and is~\cite{Ber78} 
\begin{equation}
\eta =\frac{-W^{\prime \prime }(0)}{2\pi }  \label{Eq. eta}
\end{equation}

A useful measure of how rapidly screening sets in is 
\begin{subequations}
\begin{eqnarray}
I_{C}\left( \rho \right) &=&2\pi \int_{0}^{\rho }rC(r)dr  \label{Eq. 10a} \\
&=&-1-\frac{1}{W^{\prime \prime }(0)}\left[ \frac{\left( W^{\prime }(\rho
)\right) ^{2}}{1-W^{2}(\rho )}\right] .  \label{Eq. 10b}
\end{eqnarray}
For short range screening $I\approx $ $-1$ for $\rho $ of order nearest
neighbor separations, or greater; for long range screening $I_{C}$
approaches $-1$ only when $\rho \rightarrow \infty $.

\subsection{Short range screening}

If $C(r)$ decays sufficiently rapidly with $r$, the large $r$ contribution
in the first integral in Eq. (\ref{Eq. 4}) is negligible, and the upper
limit can be extended to infinity. \ Using the screening relationship in Eq.
(\ref{Eq. 1}) we then have 
\end{subequations}
\begin{equation}
\left\langle Q^{2}\right\rangle =2\pi \eta \int_{0}^{2R}rB(r)C(r)dr.
\label{Eq. 11}
\end{equation}
In this same limit $B\left( r\right) $ in Eq. (\ref{Eq. 3b}) can be replaced
by its leading term in an expansion in powers of $r/R$, $B(r)\sim -2rR$, and
we have 
\begin{equation}
\left\langle Q^{2}\right\rangle =\frac{1}{4}\eta \Lambda _{s}P,
\label{Eq. 12}
\end{equation}
where $P=2\pi R$ is the length of the perimeter of $A$, and the screening
length $\Lambda _{s}$, which is a measure of the width of $I_{C}\left( \rho
\right) $ in Eqs. ($10$), is~\cite{FreWil98} 
\begin{equation}
\Lambda _{s}=-8\int_{0}^{\infty }r^{2}C(r)dr.  \label{Eq. LambdaScrn}
\end{equation}
Thus, the hallmark of short range screening is that for $R>>\Lambda _{s}$, $%
\left\langle Q^{2}\right\rangle $ grows with the perimeter, rather than with
the area, of $A$.

These results\ have a simple physical explanation~\cite{FreWil98}. \ Because charges
deep inside $A$ are perfectly screened they make no contribution to the net
charge $Q$, and therefore no contribution to its variance $\left\langle
Q^{2}\right\rangle $. \ Charges located within a distance of order $\Lambda
_{s}$ from the boundary of $A$, however, are imperfectly screened. \ The
number of these charges is $n\approx \eta \Lambda _{s}P$. \ For completely
unscreened charges $\left\langle Q^{2}\right\rangle \approx n$, whereas for
partially screened charges we can expect $\left\langle Q^{2}\right\rangle
\approx \alpha n$~\cite{FreWil98}, where from\ Eq. (\ref{Eq. 12}) $\alpha =\frac{1}{4}$%
.

Eqs. (\ref{Eq. 12}) and (\ref{Eq. LambdaScrn}) also quantify the notion of
short-range screening. \ The condition that $\Lambda _{s}$ is finite is
sufficient to insure that $\left\langle Q^{2}\right\rangle $ grows linearly
with $R$. \ Thus, for short range screening $C(r)$ (averaged over a period
in the case that it oscillates) must fall faster than $1/r^{3}$.

\subsubsection{Gaussian correlations}

The canonical short range correlation function which gives rise to short
range screening is the Gaussian, which we write as 
\begin{subequations}
\begin{eqnarray}
W\left( v\right) &=&\exp \left( -\kappa ^{2}a^{2}v^{2}\right) ,
\label{Eq. 14a} \\
\kappa &=&2\pi /\left( \lambda Z\right) ,  \label{Eq. 14b}
\end{eqnarray}
where $\lambda $ is the wavelength, $Z$ is the (asymptotically large)
distance between the random source and the screen on which the field $E$ is
measured, and $v$ measures radial displacements on this screen.

$W(v)$ is the normalized autocorrelation function of the far field speckle
pattern produced by a distribution of randomly phased sources with
amplitudes 
\end{subequations}
\begin{equation}
S(u)=\exp \left( -\left[ u/\left( 2a\right) \right] ^{2}\right) ,
\label{Eq. 15}
\end{equation}
where $u$ measures radial displacements in the source plane, and $a$ is a
measure of the width of the distribution \footnote{%
Such a source distribution can be obtained in practice by illuminating a
random phase screen (typically a piece of finely ground glass) with a
Gaussian laser beam.}. \ The displacement ${\bf r}$ and the displacement $%
{\bf v}$ are related by the scale factor $\kappa $ 
\begin{equation}
{\bf r}=\kappa {\bf v},  \label{Eq. 16}
\end{equation}
giving for the Gaussian 
\begin{equation}
W\left( r\right) =\exp \left( -a^{2}r^{2}\right) .  \label{Eq. 17}
\end{equation}
Then, using Eq. (\ref{Eq. LambdaScrn}) we obtain 
\begin{equation}
\Lambda _{s}=\zeta \left( 3/2\right) /\left( \sqrt{2\pi }a\right) =1.0422/a,
\label{Eq. 18}
\end{equation}
where $\zeta $ is the Riemann zeta function.

\begin{figure}
\includegraphics[width=0.7\textwidth]{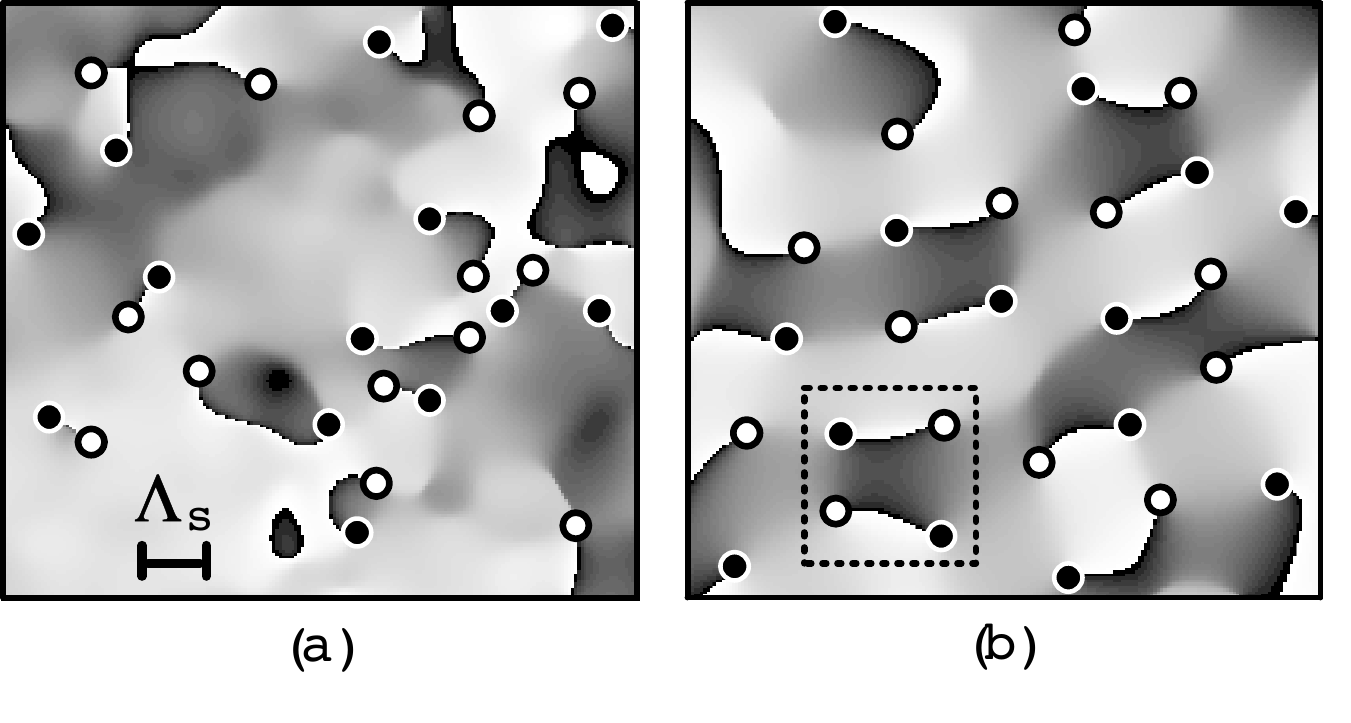}
\caption{ Singularity structure. \ Shown is the phase of the optical field
coded $0$ to $2\pi $ black to white for a field with (a) a Gaussian and (b)
a $J_{0}$ correlation function. \ Positive (negative) singularities are
shown by filled white (black) circles. \ Both fields have the same average
number density of singularities. \ (a) Gaussian correlation function. \ Here
there are $13$ positive and $13$ negative singularities. The screening
length $\Lambda _{s}$ is shown by the bar. \ Note the clustering of
singularities into small, charge-neutral groups with the spacing between
nearest neighbor singularities being of order $\Lambda _{s}$, and with large
empty regions between groups. It is this clustering that produces a
screening length $\sim 1/2$ the average spacing between singularities. \ (b) 
$J_{0}$ correlation function. \ Here there are $14$ positive and $14$
negative singularities. \ As can be seen, there is a strong tendency for the
singularities to be nearly equally spaced on a lattice-like structure in
which each nearly square unit cell, shown by the dotted rectangle, contains
two positive and two negative singularities. This structure exhibits
substantial local charge neutralization, but because of defects in the
lattice the overall screening is long range.
}
\label{fig1}
\end{figure}

The average spacing $d$ between singularities is 
\begin{equation}
d=\sqrt{1/\eta }=\sqrt{\pi }/a=1.77/a,  \label{Eq. 19}
\end{equation}
which exceeds the screening length $\Lambda _{s}$ by nearly a factor of
two! \ This apparent paradox is resolved in Fig. \ref{fig1}, where it can be seen
that the singularities tend to cluster with nearest neighbor spacings of
order $\Lambda _{s}$.

A measure of how quickly screening sets in, thereby leading to local charge
neutrality, is the rate of decay of $I_{C}\left( \rho \right) +1,$ Eqs.
(10): for $\rho =\Lambda _{s}$, $I_{C}+1=$ $0.3$, for $\rho =2\Lambda _{s}$,
it equals $0.003$, and for $\rho =3\Lambda _{s}$, $I_{C}+1$ has dropped
below $3\times 10^{-7}$.

\subsubsection{Smoothed boundaries}

In (\cite{BerDen00}, Eq. (4.48) and (4.49)) a Gauss-smoothed boundary is assumed, and a 
{\em constant}, i.e. $R$ independent, value for the charge variance, here $%
\left\langle Q^{2}\right\rangle _{const}$, is obtained from 
\begin{subequations}
\begin{eqnarray}
\left\langle Q^{2}\right\rangle _{const} &=&\frac{N}{2}\left[ 1+2\pi \eta
\int_{0}^{\infty }x\exp \left( -\frac{\pi x^{2}}{2A}\right) g_{Q}(x)dx\right]
,  \label{Eq. 55a} \\
&=&\frac{1}{4}\int_{0}^{\infty }x\frac{W^{\prime }\left( x\right) ^{2}}{%
1-W^{2}(x)}dx,  \label{Eq. 55b}
\end{eqnarray}
where $C\left( r\right) $ in \cite{BerDen00} is our $W\left( r\right) $, $g_{Q}\left(
x\right) =C\left( x\right) /\eta $ with $C\left( x\right) $ given in Eq. (%
\ref{Eq. 5}), $A$ is the area of the region of interest, here $A=\pi R^{2}$,
and $N=$ $\eta A$ is the total number of charges inside this region. \ Eq. (%
\ref{Eq. 55b}) is obtained from Eq. (\ref{Eq. 55a}) by using Eqs. (\ref{Eq.
1}), (\ref{Eq. 7}), (8), and (\ref{Eq. eta}), expanding $\exp \left( -\pi
x^{2}/\left( 2A\right) \right) $ keeping the leading term in $x$, and
integrating by parts.

For the Gaussian correlation function in Eq. (\ref{Eq. 17}), Eq. (\ref{Eq.
55b}) can be evaluated analytically, and we obtain 
\end{subequations}
\begin{equation}
\left\langle Q^{2}\right\rangle _{const}=\frac{\pi ^{2}}{48}=0.2056...
\label{Eq. 56}
\end{equation}
This result shows that $\left\langle Q^{2}\right\rangle _{const}$ is not
only independent of $R$, but also of $a$, and is therefore independent of
the charge density $\eta $ and all other system parameters! \ A question of
considerable interest not discussed in \cite{BerDen00} is whether or not this value
for $\left\langle Q^{2}\right\rangle _{const}$ is an intrinsic property of
the medium, or is it the result of a particular set of assumptions?

In \cite{BerDen00} it is stated that Eq. (\ref{Eq. 55a}) is obtained from a
calculation ``with the boundary of $A$ Gauss-smoothed to eliminate trivial
edge effects'', but the details of the calculation are not given. \ In order
to be able to more closely examine the physical content of Eq. (\ref{Eq. 55a}%
) we need a derivation of this equation. \ Here we briefly outline such a
derivation, starting with the definition of a boundary that is
``Gauss-smoothed''.

\begin{figure}
\includegraphics[width=0.5\textwidth]{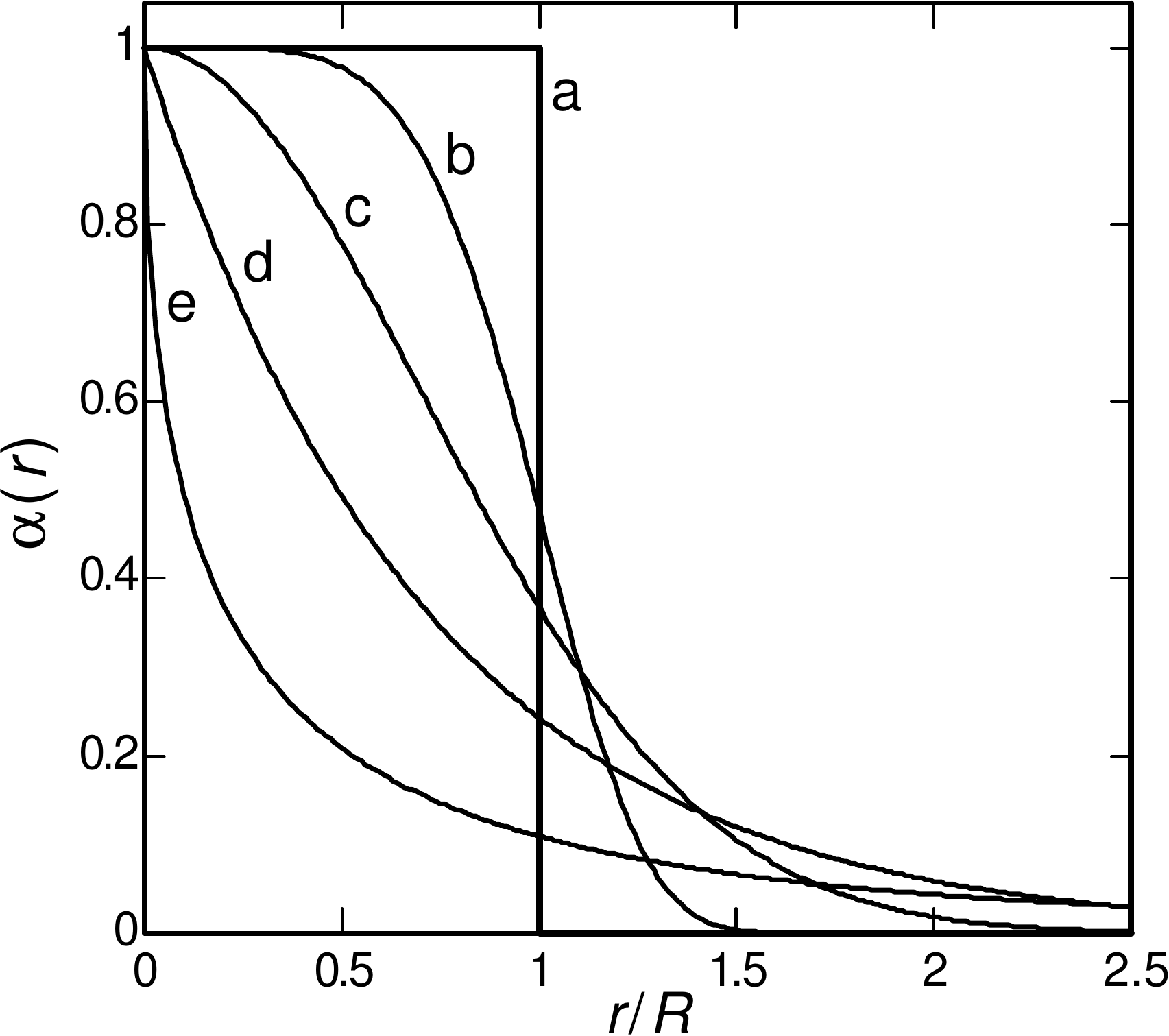}
\caption{Area functions. Shown is the normalized area function $\alpha (r)$
for different values of $p$ in Eq. (\ref{Eq. 64}).  Curve a is the disk area
function, Eq. (\ref{Eq. 58}), formally Eq. (\ref{Eq. 64}) with $p\rightarrow
\infty $. \ Curves b - e are Eq. (\ref{Eq. 64}) with the following values
for $\int $: curve b, $p=5$, a so called super Gaussian; curve c, $p=2$, a
normal Gaussian; curve d, $p=1$, a normal exponential; curve e, $p=1/2$, a
so called stretched exponential.}
\label{fig2}
\end{figure}

${\cal A}\left( x\right) $ in Eq. (3) can be obtained from an area function $\alpha
({\bf r})$ using 
\begin{equation}
{\cal A}\left( {\bf x}\right) =2\int_{0}^{\infty }rdr\int_{0}^{\pi }d\theta %
\left[ \alpha \left( {\bf r}\right) \alpha \left( {\bf r}+{\bf x}\right) %
\right] .  \label{Eq. 57}
\end{equation}
Throughout this report we have used a disk shaped area described by the area
function 
\begin{eqnarray}
\alpha ({\bf r}) &=&\alpha \left( r\right) =1,\quad 0\leq r\leq R,  \nonumber
\\
&=&0,\quad r>R.  \label{Eq. 58}
\end{eqnarray}
This function, shown in Fig. \ref{fig2} as curve (a), inserted into Eq. (\ref{Eq. 57}%
) yields 
\begin{equation}
{\cal A}\left( x\right) =4\int_{x/2}^{R}rdr\int_{0}^{\arccos \left(
x/(2r\right) }d\theta ,  \label{Eq. 59}
\end{equation}
which when evaluated yields Eq. (3).

We note that in order to maintain the same number of charges $N$ within $A$
as there are for the disk, any other form for $\alpha \left( r\right) $ must
satisfy 
\begin{equation}
2\pi \int_{0}^{\infty }r\alpha \left( r\right) dr=A=\pi R^{2}.
\label{Eq. 60}
\end{equation}

For a Gauss-smoothed boundary we have instead of Eq. (\ref{Eq. 59}), 
\begin{equation}
\alpha \left( r\right) =\exp \left( -r^{2}/R^{2}\right) =\exp \left( -\pi
r^{2}/A\right) ,  \label{Eq. 61}
\end{equation}
which is shown in Fig. 2 as curve (c). \ Inserting Eq. (\ref{Eq. 61}) into
Eq. (\ref{Eq. 57}) yields 
\begin{equation}
{\cal A}\left( x\right) =\frac{1}{2}A\exp \left( -\frac{\pi x^{2}}{2A}%
\right) .  \label{Eq. 62}
\end{equation}

A peculiarity of Eq. (\ref{Eq. 62}) is that ${\cal A}\left( 0\right) =A/2$,
as opposed to ${\cal A}\left( 0\right) =A$, the expected result, which is
obtained from Eq. (3). \ As a result of this peculiarity, in the absence of
screening the Gauss-smoothed calculation yields $\left\langle
Q^{2}\right\rangle =N/2$, instead of the expected random walk result $%
\left\langle Q^{2}\right\rangle =N$.

Generalizing Eq. (\ref{Eq. 4}) to an arbitrary area function satisfying Eq. (%
\ref{Eq. 60}), we have 
\begin{equation}
\left\langle Q^{2}\right\rangle =\eta {\cal A}\left( 0\right) +2\pi \eta
\int_{0}^{\infty }x{\cal A}\left( x\right) C\left( x\right) dx.
\label{Eq. 63}
\end{equation}
Inserting Eq. (\ref{Eq. 62}), and using $C\left( x\right) =\eta g_{Q}\left(
x\right) $ we recover Eq. (\ref{Eq. 55a}).

Now, if the physical content of $\left\langle Q^{2}\right\rangle _{const}$
were that it is an intrinsic property of the random medium, rather than
simply the result of a calculation for a particular form for $\alpha \left(
r\right) $, then other forms for $\alpha \left( r\right) $ should yield the
same result. \ In order to test this we redo the above calculation using a
boundary smoothed by a general exponential function of the form, 
\begin{equation}
\alpha \left( r\right) =\exp \left( -b_{p}\left( r/R\right) ^{p}\right) ,
\label{Eq. 64}
\end{equation}
where $0<p<\infty $ is a real, not necessarily integer, number, and in order
to satisfy Eq. (\ref{Eq. 60}), 
\begin{equation}
b_{p}=\left[ \frac{2}{p}\Gamma \left( \frac{2}{p}\right) \right] ^{\left(
p/2\right) }.  \label{Eq. 65}
\end{equation}
Examples of $\alpha \left( r\right) $ for different values of $p$ are shown
in Fig. 2 as curves (b), (d), and (e).

Inserting Eqs. (\ref{Eq. 64}) and (\ref{Eq. 65}) into (\ref{Eq. 57}), we
find 
\begin{equation}
{\cal A}\left( 0\right) =4^{-\left( 1/p\right) }\pi R^{2},  \label{Eq. 66}
\end{equation}
so that in the absence of screening $\left\langle Q^{2}\right\rangle
=4^{-\left( 1/p\right) }N$, which approaches $N$ for large $p$, and
approaches zero as $p$ approaches zero.

For the charge variance itself we have 
\begin{equation}
\left\langle Q^{2}\right\rangle _{const}=p\left( \frac{\pi ^{2}}{96}\right)
\label{Eq. 67}
\end{equation}
which reduces to the Gaussian result in Eq. (\ref{Eq. 56}) for $p=2$, but
differs from it for other values of $p$, demonstrating that $\left\langle
Q^{2}\right\rangle _{const}$ is, in fact, not an intrinsic property of the
medium, but is dependent on the particular choice of $\alpha \left( r\right) 
$.

In \cite{FreWil98} a short range autocorrelation function of the form 
\begin{equation}
W(r)=2J_{1}(ar)/\left( ar\right)  \label{Eq. 68}
\end{equation}
was used. \ For this form for arbitrary $p$ we obtain by numerical
integration 
\begin{equation}
\left\langle Q^{2}\right\rangle _{const}=p\left( 0.168077...\right) ,
\label{Eq. 69}
\end{equation}
which is again independent of $a$.

Because both Eqs. (\ref{Eq. 67}) and (\ref{Eq. 69}) are continuous functions
of $p$, a particular value for $\left\langle Q^{2}\right\rangle _{const}$
does not uniquely specify neither the field autocorrelation function nor the
boundary smoothing function. \ For example, $\left\langle Q^{2}\right\rangle
_{const}=0.2056...$ for a Gaussian autocorrelation function and a Gaussian
boundary smoothing function for which $p=2$.\ \ But the exact same value for 
$\left\langle Q^{2}\right\rangle _{const}$ is obtained using the $J_{1}$
autocorrelation function in Eqs. (\ref{Eq. 68}) and a boundary smoothing
function with $p=1.1817...$.

\subsection{Long Range Screening}

Berry and Dennis~\cite{BerDen00} have introduced the long range correlation function 
\begin{equation}
W\left( r\right) =J_{0}\left( ar\right) ,  \label{Eq. 20}
\end{equation}
where $J_{n}$ is a Bessel function of integer order $n$. \ This form for $%
W\left( r\right) $ arises from a random source distribution $S\left(
u\right) $ that takes the form of a very thin ring, 
\begin{equation}
S\left( u\right) =\delta \left( u-a\right) .  \label{Eq. 21}
\end{equation}
Practical realization of a $J_{0}$ correlation function is discussed in
Section IV.

Using Eqs. (10) with $\rho \rightarrow \infty $ the screening relationship
in Eq. (\ref{Eq. 1}) is easily verified, whereas upon inserting Eqs. (\ref
{Eq. 7}) and (\ref{Eq. Omega}) into Eq. (\ref{Eq. LambdaScrn}) it is
immediately seen that $\Lambda _{s}$ does not converge. \ These are the
conditions for long range screening.

Inserting Eqs. (\ref{Eq. 3b}), (\ref{Eq. 7}), and (\ref{Eq. Omega}) into Eq.
(\ref{Eq. 4}), and integrating by parts, we obtain 
\begin{equation}
\fbox{${\displaystyle \langle Q^{2}\rangle =\frac{1}{2\pi}
\int_{0}^{2R}\!\sqrt{4R^{2}-r^{2}}\,
\frac{(W^{\prime }(r))^{2}} { 1-W^{2}(r)}}\,dr
$}.  \label{Eq. 22}
\end{equation}
In the remainder of this report we use this highly convenient form for $%
\langle Q^{2}\rangle $ exclusively.

For the case of $J_{0}$, Eq. (\ref{Eq. 20}), we break the integral on the
R.H.S. of Eq. (\ref{Eq. 22}) into two parts: 
\begin{equation}
\langle Q^{2}\rangle ={\mathfrak I}_{1}+{\mathfrak I}_{2},  \label{Eq. 23}
\end{equation}
where 
\begin{subequations}
\begin{eqnarray}
{\mathfrak I}_{1} &=&\frac{a^{2}}{2\pi }\int_{0}^{\Delta }dr\,\sqrt{4R^{2}-r^{2}}%
\frac{J_{1}^{2}(ar)}{1-J_{0}^{2}(ar)},  \label{Eq. 24aX} \\
{\mathfrak I}_{2} &=&\frac{a^{2}}{2\pi }\int_{\Delta }^{2R}dr\,\sqrt{4R^{2}-r^{2}%
}\frac{J_{1}^{2}(ar)}{1-J_{0}^{2}(ar)},  \label{Eq. 24bX}
\end{eqnarray}
with $1/a\ll \Delta \ll R$.

In evaluating ${\mathfrak I}_{1}$ and ${\mathfrak I}_{2}$ we will need the
asymptotic forms 
\end{subequations}
\begin{equation}
\frac{J_{1}^{2}(ar)}{1-J_{0}^{2}(ar)}\approx J_{1}^{2}(ar)\approx \frac{%
2\cos ^{2}\left( ar+\pi /4\right) }{\pi ar};\quad r\gg 1/a.  \label{Eq. 25}
\end{equation}

Expanding $\sqrt{4R^{2}-r^{2}}$ and keeping only the leading $R$ term we
have for ${\mathfrak I}_{1},$%
\begin{equation}
{\mathfrak I}_{1}\approx \frac{aR}{\pi }\int_{0}^{a\Delta }dx\,\frac{J_{1}^{2}(x)%
}{1-J_{0}^{2}(x)}.  \label{Eq. 26}
\end{equation}
This form diverges as $a\Delta \rightarrow \infty $, and so we regularize it
by writing 
\begin{subequations}
\begin{eqnarray}
{\mathfrak I}_{1} &\approx &\frac{aR}{\pi }\int_{0}^{a\Delta }dx\,\left[ \frac{%
J_{1}^{2}(x)}{1-J_{0}^{2}(x)}-J_{1}^{2}(x)+J_{1}^{2}(x)\right] ,
\label{Eq. 27a} \\
&\approx &\frac{aR}{\pi }\left[ {\cal D}+\int_{0}^{a\Delta }dx\,J_{1}^{2}(x)%
\right] ,  \label{Eq. 27b}
\end{eqnarray}
where 
\end{subequations}
\begin{equation}
{\cal D}=\int_{0}^{\infty }dx\,\frac{J_{0}^{2}(x)J_{1}^{2}(x)}{1-J_{0}^{2}(x)%
}=0.5630468586...  \label{Eq. 26X}
\end{equation}
is evaluated numerically. \ Using Eq. (\ref{Eq. 25}) the remaining integral
in Eq. (\ref{Eq. 27b}) can be evaluated analytically in the limit of large $%
a\Delta $ as follows: 
\begin{subequations}
\begin{eqnarray}
\int_{0}^{a\Delta }J_{1}^{2}(x)dx &=&\lim_{\epsilon \rightarrow
0^{+}}\int_{0}^{a\Delta }dx\,x^{-\epsilon }J_{1}^{2}(x),  \label{Eq. 27aX} \\
&=&\lim_{\epsilon \rightarrow 0^{+}}\left[ \int_{0}^{\infty
}dx\,x^{-\epsilon }J_{1}^{2}(x)-\int_{a\Delta }^{\infty }dx\,x^{-\epsilon
}J_{1}^{2}(x)\right] ,  \label{Eq. 27bX} \\
&\approx &\lim_{\epsilon \rightarrow 0^{+}}\left[ \frac{\Gamma \left( \frac{%
3+\epsilon }{2}\right) }{2^{\epsilon }\Gamma (2)\Gamma \left( \frac{%
1+\epsilon }{2}\right) }{}_{2}F_{1}\left( \frac{3-\epsilon }{2},\frac{%
1-\epsilon }{2};2;1\right) -\int_{a\Delta }^{\infty }x^{-\epsilon }\frac{1}{%
\pi x}\right] ,  \label{Eq. 27cX} \\
&=&\lim_{\epsilon \rightarrow 0^{+}}\left[ \frac{\Gamma \left( \frac{%
3-\epsilon }{2}\right) }{2^{\epsilon }\Gamma (2)\Gamma \left( \frac{%
1+\epsilon }{2}\right) }\frac{\Gamma (2)\Gamma (\epsilon )}{\Gamma \left( 
\frac{1+\epsilon }{2}\right) \Gamma \left( \frac{3+\epsilon }{2}\right) }-%
\frac{1}{\epsilon \pi }(a\Delta )^{-\epsilon }\right] ,  \label{Eq. 27dX} \\
&=&\frac{\gamma +3\ln 2-2}{\pi }+\frac{1}{\pi }\ln (a\Delta ),
\label{Eq. 27eX}
\end{eqnarray}
where $\gamma =0.5772...$ is Euler's constant.

Similarly, in calculating ${\mathfrak I}_{2}$, Eq. (\ref{Eq. 24bX}), we use the
asymptotic form in Eq. (\ref{Eq. 25}) to obtain 
\end{subequations}
\begin{subequations}
\begin{eqnarray}
{\mathfrak I}_{2} &\approx &\frac{a}{2\pi ^{2}}\int_{\Delta }^{2R}dr\,\frac{%
\sqrt{4R^{2}-r^{2}}}{r},  \label{Eq. 29aX} \\
&\approx &\frac{a}{2\pi ^{2}}\left[ \sqrt{4R^{2}-x^{2}}-2R\ln \frac{4R^{2}+2R%
\sqrt{4R^{2}-x^{2}}}{x}\right] _{\Delta }^{2R},  \label{Eq. 29bX} \\
&\approx &\frac{a}{2\pi ^{2}}\left[ -2R\ln \frac{4R^{2}}{2R}-2R+2R\ln \frac{%
8R^{2}}{\Delta }\right] ,  \label{Eq. 29cX} \\
&=&\frac{aR}{\pi ^{2}}\left[ \ln \frac{4R}{\Delta }-1\right] .
\label{Eq. 29dX}
\end{eqnarray}
Combining ${\mathfrak I}_{1}$ and ${\mathfrak I}_{2}$, we obtain the central result
of this report, 
\end{subequations}
\begin{subequations}
\begin{equation}
\fbox{${\displaystyle \left\langle Q^{2}\right\rangle =\frac{aR}{\pi^2}
\left[ K+\ln \left( aR\right) \right]} $},  \label{Eq. 30bX}
\end{equation}
where 
\begin{equation}
K=\pi {\cal D}+\gamma +5\ln 2-3=2.81181544...  
\end{equation}
\end{subequations}

We note that for rather modest values $aR>17$ the $\ln \left( aR\right) $
term dominates, so that this term should be easily accessible to experiment.
\ Direct numerical integration of Eq. (\ref{Eq. 2}) reveals that Eq. (\ref
{Eq. 30bX}) is good to better than $3\%$ for $aR=2$, to better than $0.3\%$
for $aR=10$, and that thereafter the percentage error decreases as $\sim
3/(aR)$.

As for the Gaussian, a measure of how quickly (or slowly) screening sets in
is the average rate of decay of $I_{C}\left( \rho \right) +1$. \ For large $%
\rho $ this average rate is $\sim 2/\left( \pi a\rho \right) $, showing that
although screening is ultimately complete, it approaches completeness very
slowly.

\subsubsection{Smoothed boundaries}

What does boundary smoothing do to $\left\langle Q^{2}\right\rangle $ when
the correlations are long ranged?

For the long range correlation function $J_{0}\left( ar\right) $, \cite{BerDen00}
states ``we can show from (4.48) [here Eq. (\ref{Eq. 55a})] that ... $%
\left\langle Q^{2}\right\rangle \sim \sqrt{N}$'', i.e. for a circular region
of radius $R$, $\left\langle Q^{2}\right\rangle \sim R$.

For large $R$, the limit also used in \cite{BerDen00}, we find here for all
generalized exponential smoothing functions in Eq. (\ref{Eq. 64}). 
\begin{equation}
\left\langle Q^{2}\right\rangle =\Xi \left( p\right) aR.  \label{Eq. DuM_1}
\end{equation}

For the Gaussian smoothing function used in \cite{BerDen00} we obtain analytically 
\begin{equation}
\Xi \left( 2\right) =\frac{1}{4\sqrt{2\pi }}=0.09974.  \label{Eq. DuM_2}
\end{equation}

In general, however, $\Xi \left( p\right) $ must be evaluated numerically. \
Introducing scaled variables $\widetilde{x}=b_{p}^{1/p}x/R,\widetilde{r}%
=b_{p}^{1/p}r/R$, and using the asymptotic form $\Omega ^{2}\left( x\right)
\approx 1/\left( \pi x\right) $, Eq. (\ref{Eq. 25}), we have 
\begin{subequations}
\begin{eqnarray}
\Xi \left( p\right) &=&\frac{-1}{2\pi ^{2}b_{p}^{1/p}}\int_{0}^{\infty }d%
\widetilde{x}\left( \frac{d{\cal A}\left( \widetilde{x}\right) }{d\,%
\widetilde{x}}\right) ,  \label{Eq. DuM_3a} \\
{\cal A}\left( \widetilde{x}\right) &=&2\int_{0}^{\infty }\widetilde{r}d%
\widetilde{r}\int_{0}^{\pi }d\theta \exp \left( -\widetilde{r_{+}}^{p}-%
\widetilde{r_{-}}^{p}\right) ,  \label{Eq. DuM_3b} \\
\widetilde{r_{\pm }} &=&\sqrt{\widetilde{r}^{2}+\widetilde{x}^{2}/4\pm 
\widetilde{r}\widetilde{x}\cos \theta }.  \label{Eq. DuM_3c}
\end{eqnarray}
Carrying out the required numerical integrations we obtain: $\Xi \left(
4\right) =0.1615$; $\Xi \left( 6\right) =0.2052$; $\Xi \left( 8\right)
=0.2338$; and $\Xi \left( 10\right) =0.2561$.

Thus, also for long range correlations the results of boundary smoothing
change when the arbitrary smoothing function is changed.

\section{CHARGE VARIANCE IN A BOUNDED REGION AND THE ZERO CROSSING
CORRELATION FUNCTION $\Gamma $}

As indicated in the Introduction, $Q$, and therefore $\left\langle
Q^{2}\right\rangle $, can be obtained from measurements made only on the
boundary of the region of interest $A$, here a circle of radius $R$. \ The
method suggested in \cite{FreWil98}, which is amenable to calculation, is the following:
\ Signed zero crossings (ZCs) of either the real ${\cal R}$, or of the
imaginary ${\cal I}$, parts of the wave functions that cross the boundary
are counted, with each positive (negative) ZC contributing $+\frac{1}{2}
$ ($-\frac12$) to $Q$; the sign $\sigma _{{\cal R}}$ of a zero crossing of ${\cal R}$ is 
$\sigma _{{\cal R}}=$ sign$\left( -{\cal IR}_{s}\right) $, that of a ZC of $%
{\cal I}$ is $\sigma _{{\cal I}}=$ sign$\left( {\cal RI}_{s}\right) $, where 
${\cal R}_{s}=\partial {\cal R}/\partial s$, ${\cal I}_{s}=\partial {\cal I}%
/\partial s$~\cite{FreWil98}.

Writing the lineal number density of positive (negative) zero crossings of $%
{\cal R}$ or of ${\cal I}$ as $n_{+}$,($n_{-}$), their sum and difference
are 
\end{subequations}
\begin{subequations}
\begin{eqnarray}
n_{0} &=&n_{+}+n_{-},  \label{Eq. 31aX} \\
\Delta N &=&n_{+}-n_{-}.  \label{Eq. 31bX}
\end{eqnarray}
where, for circular Gaussian statistics~\cite{Ric54}
\end{subequations}
\begin{equation}
n_{0}=\frac{1}{\pi }\sqrt{-W^{\prime \prime }\left( 0\right) }.
\label{Eq. 32X}
\end{equation}

For a straight line of length $L$ oriented along say the $x$-axis~\cite{FreWil98}, 
\begin{equation}
\left\langle \left( \Delta N\right) ^{2}\right\rangle _{L}=\frac{1}{4}n_{0}L%
\left[ 1+2\int_{0}^{L}\Gamma \left( \Delta x\right) d\left( \Delta x\right) -%
\frac{2}{L}\int_{0}^{L}\left| \Delta x\right| \Gamma \left( \Delta x\right)
d\left( \Delta x\right) \right] .  \label{Eq. 33X}
\end{equation}
\ The zero crossing correlation function $\Gamma $ is 
\begin{subequations}
\begin{eqnarray}
\Gamma (\Delta x) &=&-\frac{1}{\pi ^{2}n_{0}}\frac{(1-W^{2}(\Delta
x)W^{\prime \prime }(\Delta x)+W(\Delta x)(W^{\prime }(\Delta x))^{2}}{%
(1-W^{2}(\Delta x))^{3/2}}\sin ^{-1}W(\Delta x)  \label{Eq. 34aX} \\
&=&\frac{-1}{\pi ^{2}n_{0}}\arcsin \left( W\left( \Delta x\right) \right) 
\frac{d\Omega (\Delta x)}{d\left( \Delta x\right) },  \label{Eq. 34bX}
\end{eqnarray}
where $\Omega $ is obtained from Eq. (\ref{Eq. Omega}) with $r$ replaced by $%
\Delta x$. \ $\Gamma (\Delta x)$ and $C\left( r\right) $ are compared in
Fig. \ref{fig3}.

\begin{figure}
\includegraphics[width=0.7\textwidth]{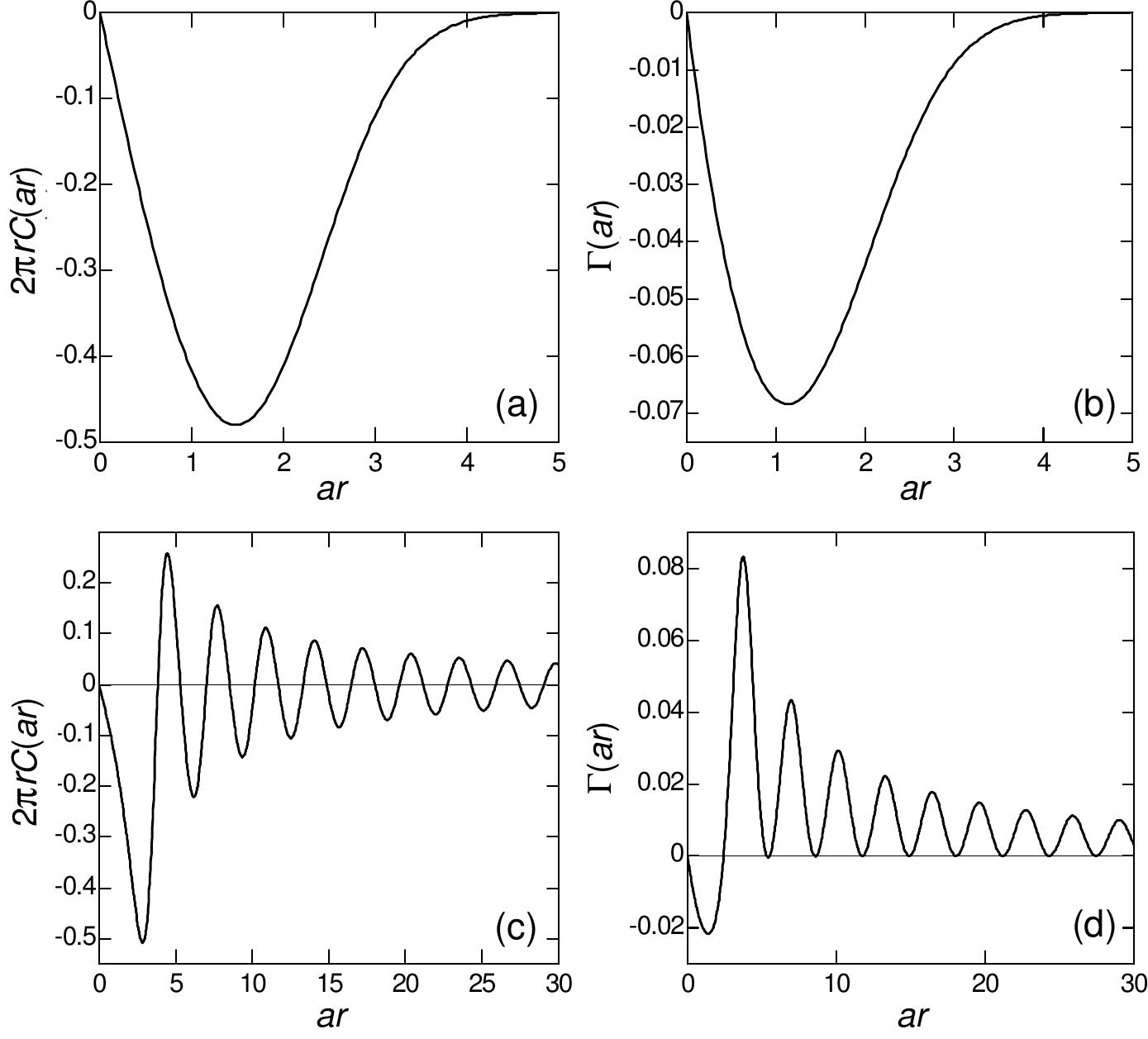}
\caption{Correlation Functions. (a,b) Short range correlations. (a) Charge
correlation function $C(ar)$ and (b) Zero crossing correlation function $%
\Gamma (ar)$ for short range Gaussian field correlator $%
W(ar)=exp(-a^{2}r^{2})$. \ (c,d) Long range correlations. \ (c) Charge
correlation function $C(ar)$ and (d) Zero crossing correlation function $%
\Gamma (ar)$ for long range field correlator $W(ar)=J_{0}(ar)$. \ Short
range correlations: $C(ar)$ in (a) is everywhere negative, indicating that
on average positive (negative) charges are surrounded by a tight cloud of
excess negative (positive) charge, leading to local charge neutralization -
this is the characteristic field structure of short range screening, see
Fig. 1(a). \ Similarly, $\Gamma (ar)$ in (b) is also everywhere negative,
indicating that a positive (negative) zero crossing on the boundary is
likely to be followed by an excess of negative (positive) zero crossings
that quickly cancel its contribution to the net charge. \ This structure
leads to a linear increase of the charge variance with boundary length. \
Long range correlations: \ Although for, say, a central positive charge the
first negative dip in $C(ar)$ in (c) integrates to an excess negative charge
of $-0.76$, indicating substantial, albeit incomplete, local charge
neutralization, much of this is offset by the following positive peak. \ As
can be seen, the net average field structure consists of rings whose weak
excess charge alternates positive/negative, with each successive negative
ring having marginally greater area, and therefore marginally greater
negative charge, than the positive ring that follows it. \ These oscillating
cancellations give rise to a slow $1/r$ approach to complete screening that
is the hallmark of long range screening. \ This weak structure is too
diffuse to be seen in Fig. 1(b). \ For the zero crossing correlation
function $\Gamma (ar)$ in (d), a positive (negative) zero crossing on the
boundary is likely to be followed by a long string of zero crossings with a
weak excess of positive (negative) signs that reflect the bulk ring
structure (the boundary runs through one of these rings). \ This structure
leads to an $R\ln R$ growth of the charge variance.
}
\label{fig3}
\end{figure}

For a circle of radius $R$, 
\end{subequations}
\begin{eqnarray}
\left\langle \left( \Delta N\right) ^{2}\right\rangle _{C} &=&\left\langle
Q^{2}\right\rangle ,  \nonumber \\
&=&\frac{1}{4}n_{0}\left( 2\pi R\right) \left[ 1+2\int_{0}^{\pi }\Gamma
\left( \Delta \theta \right) d\left( \Delta \theta \right) \right] ,
\label{Eq. 35X}
\end{eqnarray}
where $\Delta \theta $ measures the angular separation of two points on the
rim of the circle. \ This form for $\left\langle \left( \Delta N\right)
^{2}\right\rangle $ follows from Section F of \cite{FreWil98} with $x$ replaced by $%
\theta $. \ The corresponding form for $\Gamma \left( \Delta \theta \right) $
is 
\begin{equation}
\Gamma \left( \Delta \theta \right) =\frac{-1}{\pi ^{2}n_{0}^{\left( \theta
\right) }}\arcsin \left( W\right) \frac{d\Omega (\Delta \theta )}{d\left(
\Delta \theta \right) },  \label{Eq. 36X}
\end{equation}
where $n_{0}^{\left( \theta \right) }$ is the number of zero crossings per
unit of arc that cross the rim of the circle, and $\Omega (\Delta \theta )$
is obtained from Eq. (\ref{Eq. 34bX}) with $\Delta x$ replaced by $\Delta
\theta $. \ This form for $\Gamma \left( \Delta \theta \right) $ follows
from Section G of \cite{FreWil98} with $x$ replaced by $\theta $.

Using the fact that in the limit $\Delta \theta $ goes to zero the
difference between the rim (arc) of the circle $R\Delta \theta $ and the
corresponding chord$,$ here $r$, vanishes, together with $n_{0}^{\left(
\theta \right) }\left| d\theta \right| =n_{0}\left| dr\right| $, we have 
\begin{equation}
n_{0}^{\left( \theta \right) }=Rn_{0},  \label{Eq. 37X}
\end{equation}
where $n_{0}$, given in Eq. (\ref{Eq. 31aX}), is the lineal number density
of zero crossings.

In what follows we refer to Eq. (\ref{Eq. 33X}) as the {\em linear} $\Gamma $
formula, and to Eq. (\ref{Eq. 35X}) as the {\em circular} $\Gamma $ formula.

The field autocorrelation function $W\left( \Delta \theta \right) $
appearing in $\Omega (\Delta \theta )$ depends only on the straight line
separation, i.e. chord length $r$, between two points on the circle rim
separated by $\Delta \theta $, 
\begin{subequations}
\begin{eqnarray}
W\left( \Delta \theta \right) &=&W\left( r\right) ,  \label{Eq. 38aX} \\
r &=&2R\sin \left( \Delta \theta /2\right) .  \label{Eq. 38bX}
\end{eqnarray}

\subsection{Exact Equivalence}

We now proceed to show explicitly that $\left\langle Q^{2}\right\rangle $ in
Eq. \ref{Eq. 2}, which is written in terms of the charge correlation
function $C$ and the area of $A$, and $\left\langle Q^{2}\right\rangle $ in
the circular $\Gamma $ formula, Eq. (\ref{Eq. 35X}), which is written in
terms of the zero crossing correlation function $\Gamma $ and the perimeter
of $A$, are identical - as they must be.

Returning to Eq. (\ref{Eq. 22}) we write

\end{subequations}
\begin{subequations}
\begin{eqnarray}
\langle Q^{2}\rangle &=&\frac{1}{2\pi }\int_{0}^{2R}dr\,\sqrt{4R^{2}-r^{2}}%
\frac{W^{\prime }(r)}{\sqrt{1-W^{2}(r)}}\frac{d}{dr}\sin ^{-1}W(r),
\label{Eq. 39aX} \\
&=&\frac{1}{2\pi }\left\{ \pi ^{2}Rn_{o}-\int_{0}^{2R}dr\,\sin ^{-1}W(r)%
\frac{d}{dr}\left[ \sqrt{4R^{2}-r^{2}}\frac{W^{\prime }(r)}{\sqrt{1-W^{2}(r)}%
}\right] \right\} .  \label{Eq. 39bX}
\end{eqnarray}

We now compare this with the circular $\Gamma $ formula, Eq. (\ref{Eq. 35X}%
). \ Writing 
\end{subequations}
\begin{equation}
\Gamma (\Delta \theta )==-\frac{1}{\pi ^{2}n_{o}^{\theta }}\sin
^{-1}W(r)R\cos (\Delta \theta /2)\frac{d}{dr}\left[ (1-W^{2}(r))^{-1/2}R\cos
(\Delta \theta /2)\frac{d}{dr}W(r)\right] ,  \label{Eq. 40X}
\end{equation}
and using the fact that $0\leq \Delta \theta \leq \pi $, so that $\cos
\left( \Delta \theta /2\right) \geq 0=\sqrt{4R^{2}-r^{2}}/(2R)$, we have 
\begin{subequations}
\begin{eqnarray}
\langle (\Delta N)^{2}\rangle _{C} &=&\frac{n_{o}\pi R}{2}-\frac{1}{\pi }%
\int_{0}^{2R}\frac{dr}{R\cos (\Delta \theta /2)}R\cos (\Delta \theta /2)\sin
^{-1}W(r)  \nonumber \\
&&\times \frac{d}{dr}\left[ (1-W^{2}(r))^{-1/2}R\cos (\Delta \theta /2)\frac{%
dW(r)}{dr}\right] ,  \label{Eq. 41aX} \\
&=&\frac{n_{o}\pi R}{2}-\frac{1}{2\pi }\int_{0}^{2R}dr\sin ^{-1}W(r)\frac{d}{%
dr}\left[ \sqrt{4R^{2}-r^{2}}\frac{W^{\prime }(r)}{\sqrt{1-W^{2}(r)}}\right]
,  \label{Eq. 41bX}
\end{eqnarray}
{\em Q.E.D.}

\subsection{Approximate Equivalences}

It was suggested in \cite{FreWil98} that the zero crossings of ${\cal R}$ or of ${\cal %
I}$ need not necessarily be counted over the actual perimeter $P$ of ${\cal A%
}$, but rather that in an isotropic system one could perform the count over
any straight line with the same length as $P$. \ This suggestion, supported
by numerical analysis and computer simulations, was based on the notion that
for short range correlations the geometry of the line - rectangular,
circular, straight, etc. - was unimportant, as long as over most of its
length its radius of curvature was large compared to the screening length $%
\Lambda _{s}$. \ Here we demonstrate the validity of this suggestion
analytically for short range correlations, and examine its validity for long
range correlations.

We define an error parameter ${\cal E}$ by 
\end{subequations}
\begin{subequations}
\begin{eqnarray}
\langle (\Delta N)^{2}\rangle _{L} &=&\langle Q^{2}\rangle +{\cal E},
\label{Eq. 42aX} \\
{\cal E} &=&{\cal E}_{1}+{\cal E}_{2}+{\cal E}_{3},  \label{Eq. 42bX}
\end{eqnarray}
where 
\end{subequations}
\begin{equation}
{\cal E}_{1}=-\frac{n_{o}}{2}\int_{0}^{2\pi R}drr\Gamma (r),  \label{Eq. 43X}
\end{equation}
\begin{equation}
{\cal E}_{2}=n_{o}\pi R\int_{2R}^{2\pi R}dr\Gamma (r),  \label{Eq. 44X}
\end{equation}
\begin{equation}
{\cal E}_{3}=\frac{1}{2\pi }\int_{0}^{2R}dr\sin ^{-1}W(r)\frac{d}{dr}\left[
\left( \sqrt{4R^{2}-r^{2}}-2R\right) \frac{W^{\prime }(r)}{\sqrt{1-W^{2}(r)}}%
\right] .  \label{Eq. 45X}
\end{equation}

\subsubsection{Short range correlations}

For large $R$, when $W$ is short ranged $\Gamma (r)$ decays to zero well
before the upper limit of the integral on the R.H.S. of Eq. (\ref{Eq. 43X})
is reached. \ Extending this limit to $\infty $, it is easily seen that $%
{\cal E}_{1}$ is a constant that does not grow with $R$, and can therefore
be neglected; for the same reasons, ${\cal E}_{2}$ in Eq. (\ref{Eq. 44X})
can similarly be neglected. \ Expanding the square root in the R.H.S. of Eq.
(\ref{Eq. 45X}), it is easily seen that also ${\cal E}_{3}$ does not grow
with $R$, so it too can be neglected. \ Thus for short range correlations
the linear $\Gamma $ formula and the exact result for $\langle Q^{2}\rangle $
are equivalent in the limit of large $R$.

This equivalence can be demonstrated more directly. \ Returning to Eq. (\ref
{Eq. 22}), we have to leading order in $r/R$%
\begin{subequations}
\begin{eqnarray}
\langle Q^{2}\rangle &\approx &\frac{R}{\pi }\int_{0}^{\infty }dr\,\frac{%
\left( W^{\prime }(r)\right) ^{2}}{1-W^{2}(r)},  \label{Eq. 46aX} \\
&=&\frac{R}{\pi }\int_{0}^{\infty }dr\frac{W^{\prime }(r)}{\sqrt{1-W^{2}(r)}}%
\frac{d}{dr}\arcsin \left( W(r)\right) ,  \label{Eq. 46bX} \\
&=&\frac{n_{0}\pi R}{2}\left[ 1+2\int_{0}^{\infty }dr\Gamma \left( r\right) %
\right] ,  \label{Eq, 46cX}
\end{eqnarray}
which is also the leading, i.e. $R$ dependent, term in the linear $\Gamma $
formula, Eq. (\ref{Eq. 33X})$.$

We therefore conclude, in full accord with \cite{FreWil98}, that for short range
screening the linear $\Gamma $ formula can, indeed, be used to measure $%
\langle Q^{2}\rangle $.

\subsubsection{Long range correlations}

For long range correlations we need the asymptotic form of $\Gamma (r)$,
which for $J_{0}$ is 
\end{subequations}
\begin{equation}
\Gamma (r)\approx \frac{a}{\pi ^{3}n_{0}}\frac{1+\sin (2ar)}{r}.
\label{Eq. 47X}
\end{equation}

To leading order the oscillating $\sin (2ar)/r$ term makes no contribution,
and we have:

\hspace{-0.25in}for ${\cal E}_{1}$, 
\begin{subequations}
\begin{eqnarray}
{\cal E}_{1} &\approx &-\int_{0}^{2\pi R}drr\frac{a}{2\pi ^{3}r},
\label{Eq. 48aX} \\
&=&\frac{aR}{\pi ^{2}},  \label{Eq. 48bX}
\end{eqnarray}
for ${\cal E}_{2}$, 
\end{subequations}
\begin{subequations}
\begin{eqnarray}
{\cal E}_{2} &\approx &R\int_{2R}^{2\pi R}dr\frac{a}{\pi ^{2}r},
\label{Eq. 49aX} \\
&=&\frac{aR}{\pi ^{2}}\ln \pi ,  \label{Eq. 49bX}
\end{eqnarray}
and for ${\cal E}_{3},$%
\end{subequations}
\begin{subequations}
\begin{eqnarray}
{\cal E}_{3} &\approx &-\frac{1}{2\pi }\int_{0}^{2R}drW(r)W^{\prime \prime
}(r)\left( \sqrt{4R^{2}-r^{2}}-2R\right) ,  \label{Eq. 50aX} \\
&\approx &-\frac{1}{2\pi }\int_{0}^{2R}dr\left( -\frac{a}{\pi r}\right)
\left( \sqrt{4R^{2}-r^{2}}-2R\right) ,  \label{Eq. 50bX} \\
&=&-\frac{aR}{\pi ^{2}}(\ln 2-1).  \label{Eq. 50cX}
\end{eqnarray}
where in evaluating Eq. (\ref{Eq. 50aX}) we use the asymptotic form 
\end{subequations}
\begin{equation}
W(r)W^{\prime \prime }(r)\approx -\frac{2a}{\pi r}\sin ^{2}\left( ar\right) .
\label{Eq. WWpp}
\end{equation}
Assembling the pieces yields 
\begin{equation}
{\cal E}=\frac{aR}{\pi ^{2}}\ln \left( \frac{2}{\pi }\right) =-.0458aR.
\label{Eq. 51X}
\end{equation}

Thus, unlike the case of short range screening, for long range screening the
linear $\Gamma $ formula is always larger than the true result; the
percentage error, however, is not severe, being $8.8\%$ for $aR=10$, $6.3\%$
for $aR=100$, and $4.9\%$ for $aR=1000$.

$\left\langle \left( \Delta N\right) ^{2}\right\rangle _{L}$ exceeds $%
\langle Q^{2}\rangle $ because setting $L=2\pi R$ in the integrals over $%
\Gamma \left( \Delta x\right) $ in Eq. (\ref{Eq. 33X}) implies that the
correlations between zero crossings act along the arc of the circle. \ But,
as already noted in connection with the circular $\Gamma $ formula, these
correlations act only along the chords of the circle. \ However, except for
the diameter, all chords are shorter than their arcs, so that $L=2\pi R$
necessarily overestimates the correlation effects.

We now inquire as to what value of $L$ in the upper limits of the
correlation integrals appearing on the R.H.S. of Eq. (\ref{Eq. 33X})
establishes equality between $\left\langle \left( \Delta N\right)
^{2}\right\rangle _{L}$ and $\langle Q^{2}\rangle $. \ Writing $L=2pR$, we
have 
\begin{subequations}
\begin{eqnarray}
\left\langle \left( \Delta N\right) ^{2}\right\rangle _{L} &=&\langle
Q^{2}\rangle +faR,  \label{Eq. 52a} \\
f &=&\frac{\ln (p/2)+1-p/\pi }{\pi ^{2}}.  \label{Eq. 52b}
\end{eqnarray}
Setting $f=0$ and solving for $p$ we obtain 
\end{subequations}
\begin{equation}
p=-\pi \text{LW}\left( -\frac{2}{\pi e}\right) =1.01701687...  \label{Eq. 53}
\end{equation}
where LW is the LambertW function~\cite{CorGon96}, and $e=2.718...$ is the base of
the natural logarithm. \ Eq. (\ref{Eq. 53}) suggests that as a physically
attractive approximation $L$ should be set equal to the longest chord, $2R$,
rather than to the longest arc, $2\pi R$. \ Indeed, inserting $p=1$ into $f$
yields the excellent approximation ${\cal E}=-0.00116aR$ - a forty-fold
improvement over Eq. (\ref{Eq. 51X}).

\section{PRACTICAL REALIZATION OF J$_{0}$}

As discussed in Section IIc, the long range correlation function $%
W_{J_{0}}\left( r\right) =J_{0}\left( ar\right) $ arises from a source $%
S\left( u\right) $ that takes the form of a ring of zero width, Eq. (\ref
{Eq. 21}). \ Here we consider a ring of mean radius $a$ and finite width $b$
as a practical realization, $W_{pJ_{0}}\left( r\right) $, of $%
W_{J_{0}}\left( r\right) $.

We obtain $W_{pJ_{0}}\left( r\right) $ from the Van Cittert-Zernike theorem~\cite{Goo85} as 
\begin{subequations}
\begin{eqnarray}
W_{pJ_{0}}\left( r\right) &=&\frac{1}{abr}\left[ \left( a+b/2\right)
J_{1}\left( ar+br/2\right) -\left( a-b/2\right) J_{1}\left( ar-br/2\right) %
\right] ,  \label{Eq. 54a} \\
&\approx &J_{0}\left( ar\right) \left[ 1-\frac{1}{6}\left( \frac{b}{2a}%
\right) ^{2}\left( ar\right) ^{2}\right] -\frac{1}{6}\left( \frac{b}{2a}%
\right) ^{2}\left( ar\right) J_{1}\left( ar\right) +{\cal O}\left( \left( 
\frac{b}{2a}\right) ^{4}\right) .  \label{Eq. 54b}
\end{eqnarray}
$W_{pJ_{0}}\left( r\right) $ closely matches $W_{J_{0}}\left( r\right) $
over the region $0\leq ar\leq a/b$, see Fig.  \ref{fig4}(a). \ We can therefore
expect for $0\leq ar\leq {\frac12} a/b$, and in fact do obtain, see Figs. \ref{fig4}(b,c), a close match between the
result obtained using $W_{pJ_{0}}$ in Eqs. (\ref{Eq. 7}) and (\ref{Eq. Omega}%
) and integrating Eq. (\ref{Eq. 22}) numerically, and that obtained using $%
W_{J_{0}}$.

\begin{figure}
\includegraphics[width=0.7\textwidth]{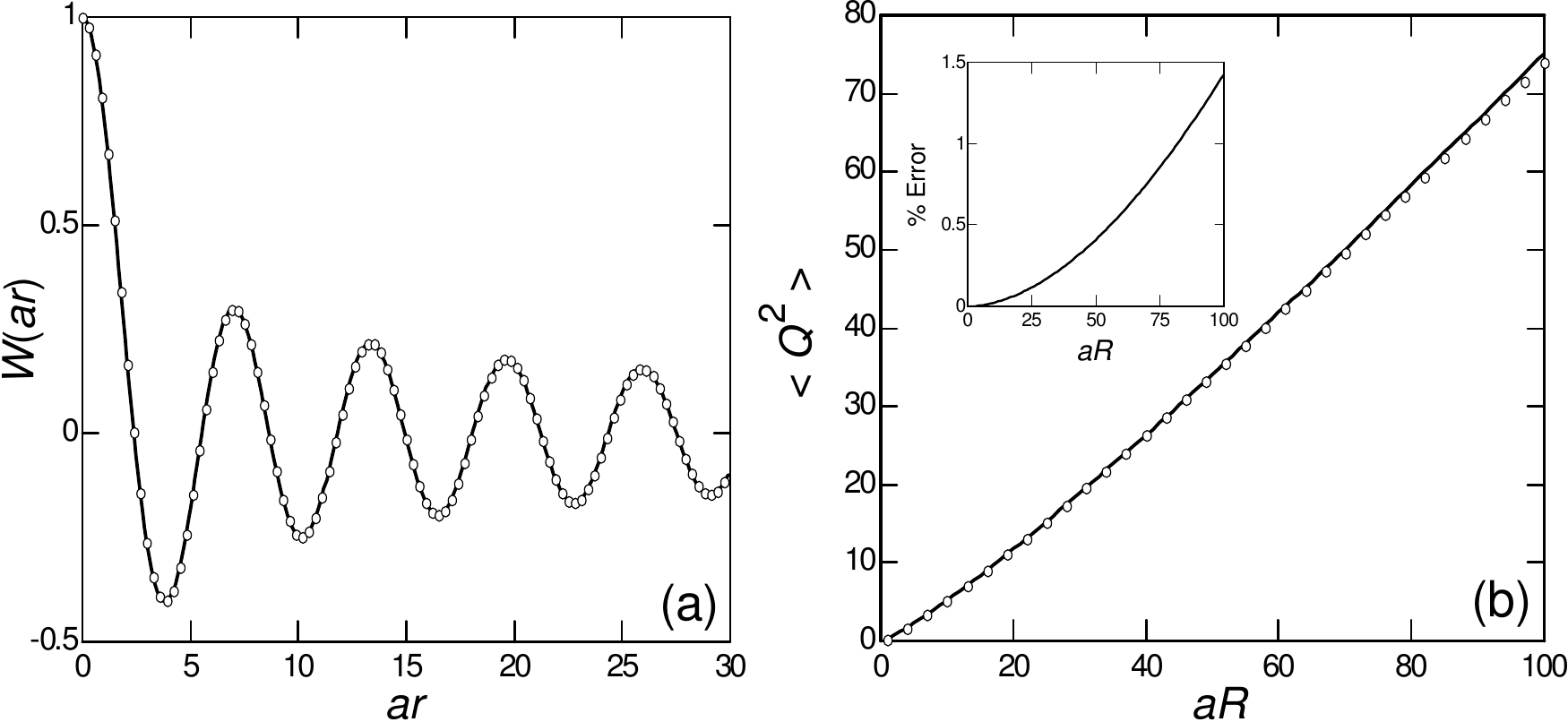}
\caption{Practical realization of $J_{0}$. (a) Correlation functions. \
Solid curve $W(ar)=J_{0}(ar)$, small circles $W(ar)$ equal to the practical
realization of $J_{0}$, Eq. (\ref{Eq. 54a}); in both cases $a=1$ and $b=0.01$%
. \ The agreement between the two forms is excellent out to $ar=100$. \ (b) $%
\left\langle Q^{2}\right\rangle $ vs. $aR$. \ Solid curve the exact result,
Eq. (\ref{Eq. 22}) with $J_{0}$ correlation function, small circles Eq. (\ref
{Eq. 22}) with practical realization of $J_{0}$. The inset shows the
percentage error obtained using the practical realization. \ This error is
less than $0.5\%$ for $aR=50$, and less than $1.5\%$ for $aR=100$.
}
\label{fig4}
\end{figure}

A ratio $b/a=100$ appears to be experimentally practical, so that the long
range screening results presented here should be amenable to experimentation.

\section{DISCUSSION}

\subsection{Smoothed boundaries}

What does smoothing the boundary of the wavefield mean both physically and
mathematically? \ Close examination of the derivation of Eq. (\ref{Eq. 55a})
reveals that {\em all} boundary smoothing functions act as filters that
smoothly reduce to zero {\em not} the field amplitude as one recedes from
the central point, but rather the {\em number density of singularities}.

How can such a fall-off be achieved?

A physically meaningful possibility is to devise a source function $S$ that
leads to a wavefield with an intrinsic singularity density that falls
smoothly to zero from some central point. \ However, not only is there no
suitable source function known, even in principle, but also such a
wavefield, if it could be created, would violate a fundamental assumption in
the derivation of the charge correlation function $C\left( r\right) $ and in
the calculation of $\left\langle Q^{2}\right\rangle $ that is used in \cite{BerDen00}
as well as here, namely that the system is stationary, i.e. that on average
it is the same everywhere, that there is no central point. \ Thus, for such
a system the results given in \cite{BerDen00}, and our extension of them, need not
necessarily apply.

So, if an intrinsic average number density of singularities that is
independent of position is needed, then how can the fall-off required by the
smooth boundary in \cite{BerDen00} be obtained? \ The only possibility that comes to
mind is the following: \ Some arbitrary point, the central point, is
surrounded by a small circle and a series of narrow concentric rings. \
Suitable fractions of the singularities in these rings are then discarded to
create a singularity density profile that approximates, say, the Gaussian in
\cite{BerDen00}, or any other desired smoothing function.

Of course, an experimentalist would object that not only is such a procedure
entirely arbitrary, but also that no good could ever come from throwing away
data. \ And, as shown above, she would be right, because the cost of
boundary smoothing for short range correlations is a loss of all wavefield
information. \ Similarly, for long range correlations the magnitude of the
calculated fluctuations change with changes in smoothing function, showing
that the coefficient $\Xi $ in Eq. (\ref{Eq. DuM_1}) cannot be interpreted
as a screening length, as is done in Eq. (\ref{Eq. 12}).

\subsection{Deviations From the Linear Law}

\subsubsection{Small $R$ for any correlation function}

Surprisingly, perhaps, for sufficiently small $R$, $\left\langle
Q^{2}\right\rangle \sim R^{2}$ for all field correlation functions for which 
$W^{\prime }\left( 0\right) =0$; this includes not only the Gaussian in Eq. (%
\ref{Eq. 17}), $J_{0}$ in Eq. (\ref{Eq. 20}), the practical approximation to 
$J_{0}$ in Eq. (\ref{Eq. 54a}), but also every other physically realizable
short and long range correlation function. \ $W^{\prime }\left( 0\right) =0$
follows from expansion of the Hankel transform in the Van Cittert-Zernike
theorem~\cite{Goo85}, and the fact that all physically realizable $S$ are bounded.
\ The proof of this universal $R^{2}$ law for $R<1/a$ is as follows: \
Returning to Eq. (\ref{Eq. 22}) we have for $ar\ll 1$, 
\end{subequations}
\begin{subequations}
\begin{eqnarray}
\left\langle Q^{2}\right\rangle &=&\eta \int_{0}^{2R}dr\sqrt{4R^{2}-r^{2}},
\label{Eq. 70a} \\
&=&\eta \pi R^{2},  \label{Eq. 70b} \\
&=&N.  \label{Eq. 70c}
\end{eqnarray}
The physical meaning of this result is that for sufficiently small $A=\pi
R^{2}$ there can no be screening because the probability of finding the
required negative charge within $A$ is vanishingly small.

\subsubsection{Large $R$ for long range correlation functions that decay
slower than $J_{0}$}

({\em i}) If one had a form for $W\left( r\right) $ for which $W^{\prime
}(r) $ decays asymptotically as $r^{-\beta }$ with $\beta <\frac12$,
 then 
\end{subequations}
\begin{equation}
\left\langle Q^{2}\right\rangle \approx \frac{\Gamma \left( {\frac12}
-\beta \right) }{2^{2\beta +1}\sqrt{\pi }\,\Gamma \left( 2-\beta \right) }%
R^{2-2\beta }.  \label{Eq. 71}
\end{equation}

({\em ii}) For $\beta ={\frac12}$, $\left\langle Q^{2}\right\rangle \approx R\ln R$, 
Eq. (\ref{Eq. 30bX}).
\ 

({\em iii}) For $\beta > {\frac12}$, $\left\langle Q^{2}\right\rangle \approx R$,
 Eq. (\ref{Eq. 12}).

The proofs of assertions ({\em i}) and ({\em iii}) are as follows:

For $\beta > {\frac12}$,
 Eq. (\ref{Eq. LambdaScrn}) converges, and therefore Eq. (\ref{Eq. 12})
holds.

For $\beta < {\frac12}$ we rewrite Eq. (\ref{Eq. 22}) as 
\begin{subequations}
\begin{eqnarray}
\left\langle Q^{2}\right\rangle &=&{\cal F}_{1}+{\cal F}_{2},
\label{Eq. 71a} \\
{\cal F}_{1} &=&\frac{1}{2\pi }\int_{0}^{\Delta }dr\sqrt{4R^{2}-r^{2}}\Omega
^{2}\left( r\right) ,  \label{Eq. 71b} \\
{\cal F}_{2} &=&\frac{1}{2\pi }\int_{\Delta }^{2R}dr\sqrt{4R^{2}-r^{2}}%
\Omega ^{2}\left( r\right) ,  \label{Eq. 71c}
\end{eqnarray}
where $\Omega $ is given in Eq. (8), and $1\ll a\Delta \ll aR$. \ For ${\cal %
F}_{1}$ we have ${\cal F}_{1}\approx \frac{R}{2}\int_{0}^{\Delta }dr\Omega
^{2}\left( r\right) $, and since this is subdominant, depending only
linearly on $R$, we discard it without further discussion. \ Inserting the
asymptotic form $\Omega ^{2}\left( r\right) \approx 1/r^{2\beta }$ into $%
{\cal F}_{2}$, and writing $s=r/\left( 2R\right) $, we have 
\end{subequations}
\begin{equation}
{\cal F}_{2}=\frac{R^{2-2\beta }}{2^{2\beta -1}\pi }\int_{\Delta /\left(
2R\right) }^{1}ds\sqrt{1-s^{2}}/s^{2\beta }.  \label{Eq. 72}
\end{equation}
Passing to the limit $\Delta /\left( 2R\right) \rightarrow 0$ we obtain Eq. (%
\ref{Eq. 71}).

\subsubsection{Physically realizable fields}

How small can $\beta $ be in practice? \ The answer is that asymptotically,
i.e. for sufficiently large $R$, $\beta \geq 3/2$. \ This is demonstrated
below using the Van Cittert-Zernike theorem~\cite{Goo85}  and the fact that all
physically realizable sources $S\left( u\right) $ must be: ({\em i})
nonsingular, a requirement that excludes, inter alia, delta functions; ({\em %
ii}) positive definite; and ({\em iii}) strictly bounded, i.e. $S\left(
u>u_{max}\right) =0$.

In physically realizable sources whose finite extent is defined by a mask
where the intensity falls discontinuously from some non-negligible value to
zero at the mask edge $u=u_{max}$, we have for large $r$%
\begin{equation}
W(r)\approx \int^{u_{max}}uS(u)J_{0}(ru)\,du\approx -\sqrt{\frac{2}{\pi }}%
(u_{max})^{1/2}S(u_{max})\cos (ru_{max}+\pi /4)/r^{3/2}\text{,}
\label{Eq. 73}
\end{equation}
so that both $W(r)$ and $W^{\prime }(r)$ decay as $r^{-3/2}$, i.e. $\beta
=3/2$.

If $S(u_{max})$ is extremely small the $r^{-3/2}$ decay of $W^{\prime }$
sets in at values of $r$ that may be so large as to be unmeasurable in
practice. \ In that case, in practice one observes $\beta \geq 4.$ \ To show
this we set $u_{max}$ equal to infinity and write 
\begin{subequations}
\begin{eqnarray}
W(r) &=&{\cal W}_{1}+{\cal W}_{2},  \label{Eq. 74a} \\
{\cal W}_{1} &=&\int_{0}^{\Delta }uS(u)J_{0}(ru)\,du,  \label{Eq. 74b} \\
{\cal W}_{2} &=&\int_{\Delta }^{\infty }uS(u)J_{0}(ru)\,du,  \label{Eq. 74c}
\end{eqnarray}
where $1\gg \Delta \gg 1/r$. \ In ${\cal W}_{1}$we expand $S(u)\approx
S(0)+uS^{\prime }(0)$ and obtain for large $r$%
\end{subequations}
\begin{eqnarray}
{\cal W}_{1} &\approx &S(0)\left[ -\sqrt{\frac{2}{\pi }}\frac{\Delta
^{1/2}\cos (r\Delta +\pi /4)}{r^{3/2}}+\frac{3}{8}\sqrt{\frac{2}{\pi }}\frac{%
\sin (r\Delta +\pi /4)}{\Delta ^{1/2}r^{5/2}}\right]  \nonumber \\
&&+S^{\prime }(0)\left[ -\sqrt{\frac{2}{\pi }}\frac{\Delta ^{3/2}\cos
(r\Delta +\pi /4)}{r^{3/2}}+\frac{11}{8}\sqrt{\frac{2}{\pi }}\frac{\Delta
^{1/2}\sin (r\Delta +\pi /4)}{r^{5/2}}-\frac{1}{r^{3}}\right] .
\label{Eq. 75}
\end{eqnarray}

Using the large argument expansion of $J_{0}$ in ${\cal W}_{2}$ and
integrating by parts we have 
\begin{subequations}
\begin{eqnarray}
{\cal W}_{2} &\approx &\sqrt{\frac{2}{\pi }}\int_{\Delta }^{\infty }uS(u)%
\left[ \frac{\sin (ru+\pi /4)}{\sqrt{ru}}-\frac{1}{8}\frac{\cos (ru+\pi /4)}{%
(ru)^{3/2}}\right] \,du  \nonumber \\
&\approx &\sqrt{\frac{2}{\pi }}\frac{\cos (r\Delta +\pi /4)}{r^{3/2}}\left[
S(0)+\Delta S^{\prime }(0)\right] \Delta ^{1/2}  \label{Eq. 76a} \\
&&-\frac{\sin (r\Delta +\pi /4)}{r^{5/2}}\left( \frac{S(0)+\Delta S^{\prime
}(0)}{2\Delta ^{1/2}}+S^{\prime }(0)\Delta ^{1/2}\right)  \nonumber \\
&&+\frac{1}{8}\frac{\sin (r\Delta +\pi /4)}{r^{5/2}}\left[ S(0)+\Delta
S^{\prime }(0)\right] \Delta ^{-1/2}  \label{Eq. 76b} \\
&=&S(0)\sqrt{\frac{2}{\pi }}\left[ \cos (r\Delta +\pi /4)r^{-3/2}\Delta
^{1/2}+\frac{3}{8}\sin (r\Delta +\pi /4)r^{-5/2}\Delta ^{-1/2}\right] 
\nonumber \\
&&+S^{\prime }(0)\sqrt{\frac{2}{\pi }}\left[ \cos (r\Delta +\pi
/4)r^{-3/2}\Delta ^{3/2}-\frac{11}{8}\sin (r\Delta +\pi /4)r^{-5/2}\Delta
^{1/2}\right]  \label{Eq. 76c}
\end{eqnarray}

Summing ${\cal W}_{1}$ and ${\cal W}_{2}$ yields 
\end{subequations}
\begin{equation}
W(r)\approx -S^{\prime }(0)/r^{3},  \label{Eq. 77}
\end{equation}
i.e. $\beta =4$. \ If, in addition, $S^{\prime }(0)=0$, the decay of $W(r)$
is faster yet. \ For example, when all odd derivatives of $S(u)$ vanish at
the origin, as happens for a Gaussian $S\left( u\right) $, $W(r)$ decays
exponentially fast.

Thus, for all physically realizable optical fields $\left\langle
Q^{2}\right\rangle $ is bounded at its endpoints by two universal laws: for
small $R$, $\left\langle Q^{2}\right\rangle \sim R^{2}$; for large $R$, $%
\left\langle Q^{2}\right\rangle \sim R$. \ This behavior is illustrated in
Fig. 5 for the practical $J_{0}$ autocorrelation function in Section IV.

\begin{figure}
\includegraphics[width=0.5\textwidth]{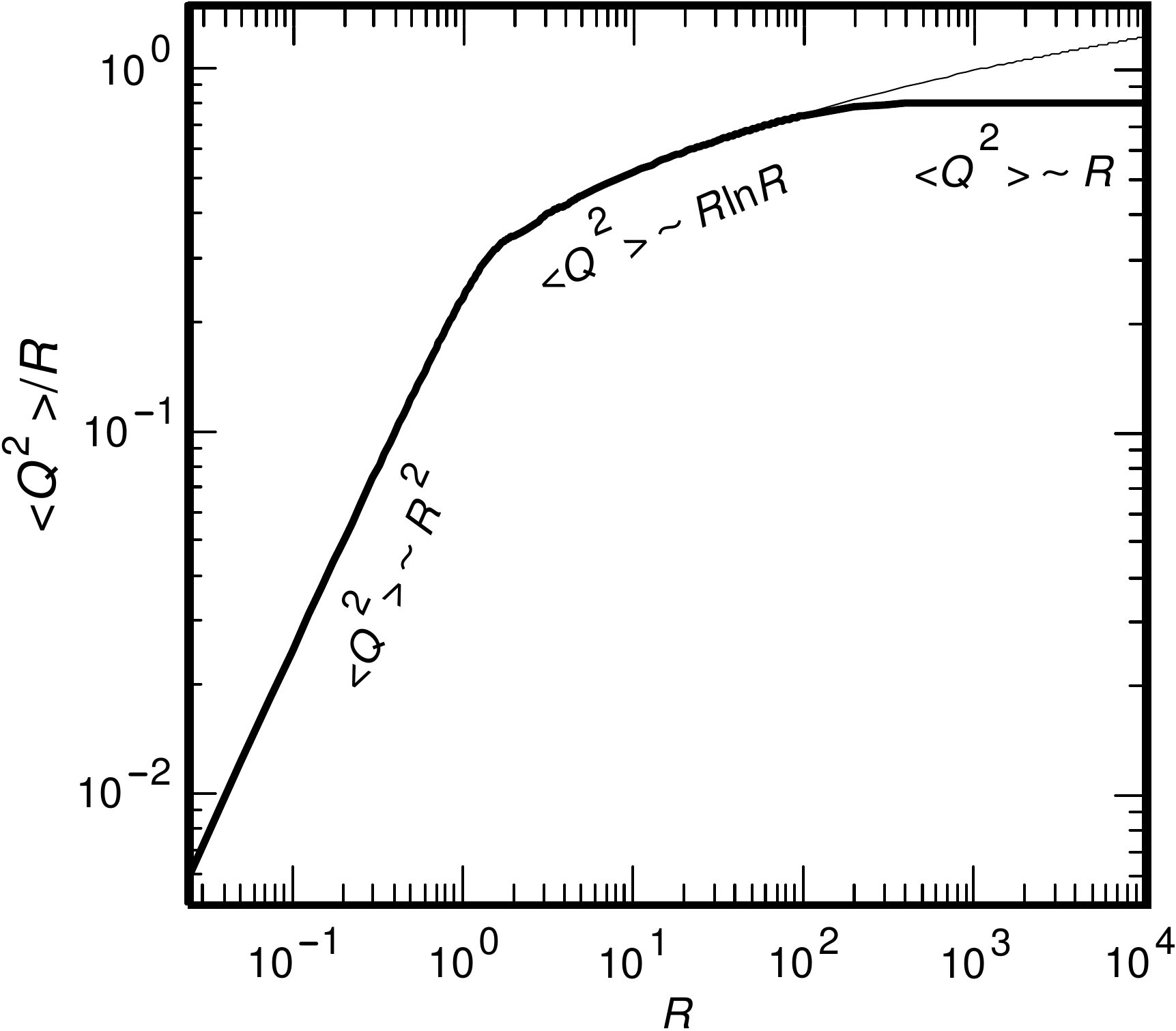}
\caption{Endpoint laws. \ Thick curve, practical $J_{0}$, Eq. (\ref{Eq. 54a}%
) with $a=1,b=0.01$; thin curve, exact $J_{0}$, Eq. (\ref{Eq. 20}) with $a=1$%
. \ Both functions obey the universal small $R$ law, $\left\langle
Q^{2}\right\rangle \sim R^{2}$, Eq. ($75$), as they must, but, as expected,
only the physically realizable practical $J_{0}$ obeys the large $R$ law, $%
\left\langle Q^{2}\right\rangle \sim R$, Eq. (\ref{Eq. 12}).}
\label{fig5}
\end{figure}

\section{SUMMARY}

Under the assumption of circular Gaussian statistics, the topological
charge variance $\left\langle Q^{2}\right\rangle $ of vortices in scalar
fields~\cite{NyeBer74} and C points in vector fields~\cite{Ber78} was analyzed for a circular
region of area $A$ and radius $R$ for both short and long range correlation
functions.

({\em i}) It was shown, Eq. ($10$), that when the autocorrelation function $%
W\left( r\right) $ together with the first derivative $W^{\prime }\left(
r\right) $ go to zero at infinity, Eq. (\ref{Eq. 1}) holds and there is
screening.

({\em ii}) For short range correlations $\left\langle Q^{2}\right\rangle $
grows linearly with $R$, Eq.  (\ref{Eq. 12}): due to screening, inside $A$
there is local charge neutralization, and only partially screened charges
near the boundary contribute to the fluctuations. \ Very recent experimental
measurements of $\left\langle Q^{2}\right\rangle $~\cite{EgoSos07} have used Eq. (\ref
{Eq. 12}) to obtain the screening length $\Lambda _{s}$, a fundamental
wavefield parameter. \ Comparison of $\Lambda _{s}$ with the mean spacing
between charges led to the conclusion that the charges form small clusters -
a conclusion verified by direct imaging of the wavefield, Fig. 1.

({\em iii}) Boundary smoothing~\cite{BerDen00} was considered for short rang
screening, and it was shown that such smoothing does not yield useful
results, Eq. (\ref{Eq. 67}).

({\em iv}) For long range correlation functions $\left\langle
Q^{2}\right\rangle $ grows faster than $R$, and it is not possible to define
a screening length. \ For a $J_{0}$ correlation function $\left\langle
Q^{2}\right\rangle \sim R\ln R$, Eq. (\ref{Eq. 30bX}). \ \ For correlation
functions that decay more slowly than $J_{0}$, $\left\langle
Q^{2}\right\rangle \sim R^{p}$, where $1<p<2$.

({\em v}) Although a $J_{0}$ correlation function is not attainable in
practice, it was shown that one can generate an excellent approximation
valid over an arbitrarily large, but finite range of $R$, Eq. (\ref{Eq. 54a}%
); this approximation was shown to yield results for $\left\langle
Q^{2}\right\rangle $ that are in close agreement with those for $J_{0}$,
Fig. 4.

({\em vi}) $\left\langle Q^{2}\right\rangle $ can be calculated using either
the charge correlation function $C\left( r\right) $, Eqs. (\ref{Eq. 7}) and (%
$8$), or the zero crossing correlation function $\Gamma \left( r\right) $,
Eq. ($53$). \ An exact calculation showed that, as expected, these two
seemingly different methods yield the same result for both short and long
range screening.

({\em vii}) For short range screening it was shown that it is also possible
to obtain $\left\langle Q^{2}\right\rangle $ from zero crossing measurements
made along any straight line of length $P=2\pi R$, rather than only along
the circular perimeter of $A$. \ For long range screening, however, it was
shown that this useful simplification no longer holds.

({\em viii}) \ It was also shown that for every physically realizable
wavefield, $\left\langle Q^{2}\right\rangle \sim R^{2}$ for sufficiently
small $R,$ and that for sufficiently large $R$, $\left\langle
Q^{2}\right\rangle \sim R$.

\acknowledgments

D.A.K. acknowledges the support of the Israel Science Foundation.

\end{document}